\documentclass[12pt,preprint]{aastex}

\begin{document}

\def\chone   {{$3.6\mu$m}}
\def\chtwo   {{$4.5\mu$m}}
\def\chthree {{$5.8\mu$m}}
\def\chfour  {{$8.0\mu$m}}
\def\zband   {{$z_{850}$-band}}
\def\iband   {{$i_{775}$-band}}
\def\vband   {{$V_{606}$-band}}
\def\bband   {{$B_{435}$-band}}
\def\jband   {{$J_{110}$-band}}
\def\hband   {{$H_{160}$-band}}
\def\ujy     {{$\mu Jy$}}

\title{High-Redshift Extremely Red Objects in the HST Ultra Deep Field Revealed by the GOODS IRAC Observations}

\author{Haojing Yan\altaffilmark{1},
Mark Dickinson\altaffilmark{4},
Peter R. M. Eisenhardt\altaffilmark{2},
Henry C. Ferguson\altaffilmark{3},
Norman A. Grogin\altaffilmark{5},
Maurizio Paolillo\altaffilmark{3,6},
Ranga-Ram Chary\altaffilmark{1},
Stefano Casertano\altaffilmark{3},
Daniel Stern\altaffilmark{2},
William T. Reach\altaffilmark{1},
Leonidas A. Moustakas\altaffilmark{3},
S. Michael Fall\altaffilmark{3}
}

\altaffiltext{1} {Spitzer Science Center, California Institute of Technology,
MS 100-22, Pasadena, CA 91125; yhj@ipac.caltech.edu}
\altaffiltext{2} {Jet Propulsion Laboratory, 4800 Oak Grove Dr., MS 169-327,
Pasadena, CA 91109}
\altaffiltext{3} {Space Telescope Science Institute, 3700 San Martin Dr.,
Baltimore, MD 21218}
\altaffiltext{4} {National Optical Astronomy Observatory, 950 N. Cherry St.,
Tucson, AZ 85719}
\altaffiltext{5} {Dept. Physics \& Astronomy, Johns Hopkins University, 3400 N.
Charles St., Baltimore, MD 21218}
\altaffiltext{6} {Dip.di Scienze Fisiche, Universit\`{a} Federico II, via Cintia 6, 80126 Napoli, Italy}

\begin{abstract}

   Using early data from the Infrared Array Camera (IRAC) on the {\it Spitzer
Space Telescope}, taken for the Great Observatories Origins Deep Survey (GOODS),
we identify and study objects that are well-detected at 3.6$\mu$m, but are
very faint (and in some cases, invisible) in the Hubble Ultra Deep Field (HUDF)
ACS and NICMOS images and in very deep VLT $K_s$-band imaging. We select a 
sample of 17 objects with $f_\nu(3.6\mu m)/f_\nu(z_{850})>20$. 
The analysis of their spectral energy-distributions (SEDs)
from 0.4 to 8.0$\mu$m shows that the majority of these objects cannot be
satisfactorily explained without a well-evolved stellar population. 
We find that most of them can be well fitted by a simple
two-component model, where the primary component represents a massive, old
population that dominates the strong IR emission, while the secondary component
represents a low amplitude, on-going star-formation process that accounts for
the weak optical fluxes. Their estimated photometric redshifts ($z_p$) range
from 1.6 to 2.9, with the median at $z_p=2.4$. For the simple star formation
histories considered here, their corresponding stellar masses range from 0.1 to 
$1.6\times 10^{11} M_\odot$ for a Chabrier Initial Mass Function (IMF).
Their median rest-frame $K_s$-band absolute 
magnitude is $-22.9$ mag in AB system, or $1.5 \times L^\ast(K)$ for
present--day elliptical galaxies. In the scenario of pure luminosity evolution,
such objects may be direct progenitors for at least 14 to 51\% of the local
population of early--type galaxies. Due to the small cosmic volume of the HUDF,
however, this simple estimate could be affected by other effects such as
cosmic variance and the strong clustering of massive galaxies. A full analysis
of the entire GOODS area is now underway to assess such effects.

\end{abstract}


\section{Introduction}

   The {\it Spitzer Space Telescope}, the fourth and last of NASA's Great
Observatories, provides order-of-magnitude improvements in capabilities of
studying the infrared sky over a wide wavelength coverage (Werner et al. 2004).
IRAC, one of the three instruments on-board {\it Spitzer}, is a four-channel
camera that simultaneously takes images at 3.6, 4.5, 5.8, and 8.0$\mu$m 
(Fazio et al. 2004). Extremely deep observations with IRAC are a key component
of the GOODS {\it Spitzer} Legacy program. By the time of writing, the first
epoch of GOODS
IRAC observations has been finished. With a typical exposure time of 23 hours
per pixel, these data have already imaged the 3.6--8.0$\mu$m sky to an 
unprecedented depth. About 1/3 of the GOODS area, including the Hubble Ultra
Deep Field (HUDF \footnote{See http://www.stsci.edu/hst/udf}; PI. S. Beckwith),
has been observed in all four IRAC channels.

   A subject of immediate interest is whether these IRAC data have revealed
any objects with unusual properties at these largely unexplored depths and
wavelengths. In this paper, we discuss a particular population that are
bright in all IRAC channels, but are extremely faint or even invisible at
optical wavelengths. The red colors of these objects are reminiscent of those
for the so-called ``Extremely Red Objects'' (EROs; e.g., Elston et al. 1988;
McCarthy et al. 1992; Hu \& Ridgway 1994; Thompson et al. 1999;
Yan et al. 2000; Scodeggio \& Silva 2000; Daddi et al. 2000), which are
commonly selected based on  $R-K$ or $I-K$ photometry.
We will therefore refer to our objects as IRAC-selected Extremely Red Objects
(IEROs), and will discuss their possible connection to conventional EROs in
this paper. To better constrain their optical fluxes, we concentrate our
discussion on the area defined by the HUDF, where the deepest optical data are
available. Throughout this paper, we adopt the following cosmological
parameters:
$\Omega_M=0.27$, $\Omega_\Lambda=0.73$, and $H_0=71$ km$\,$s$^{-1}\,$Mpc$^{-1}$.
All magnitudes are in the AB system unless specified otherwise.

\section{Data, Photometry and Sample Definition}

   The data from the first epoch of GOODS IRAC observations on the CDF-S are
described by Dickinson et al. (2004, in preparation). The nominal exposure
time per pixel is about 23.18 hours in each channel. The southern 2/3 of the
entire field has been covered by \chone\ and \chthree\ channels, while the
northern 2/3 has been covered by \chtwo\ and \chfour\ channels. The middle
1/3 of the field, which includes the HUDF, has been observed in all four IRAC
channels. The images were ``drizzle''-combined (Fruchter \& Hook 2002), and the
pixel scale of the final mosaics is $0.6''$, or approximately half of the
native IRAC pixel size. For an isolated point source, the formal detection
limits ($S/N=5$) from background shot noise in regions with full exposure time
range from 0.11 (in \chone) to 1.66 $\mu$Jy (in \chfour). In practice, crowding
and confusion influence the detection limits. As discussed later, we will 
restrict ourselves to objects which are reasonably well isolated.

   We used SExtractor (Bertin \& Arnouts 1996) in double-image mode to perform
matched-aperture photometry. We detected objects in a weighted average of the
\chone\ and \chtwo\ images in order to provide a single catalog that covers
the entire GOODS area with reasonable uniformity. A $5\times 5$ pixel Gaussian
convolving kernel with a FWHM of $1.8''$ was used for detection. We required a
detected object have a minimum connected area of two pixels (in the convolved
image) that were 1.5 $\sigma$ above the background. We adopted the photometric
calibration constants provided in the image headers generated by the Spitzer
Science Center IRAC data processing pipeline. The ``MAG\_AUTO'' option was used
throughout. Photometric errors were estimated using realistic noise maps 
generated as part of the data reduction process, but these include background
photon noise only. Crowding and source blending will generally increase the
photometric uncertainties, although in this paper we will limit ourselves to
reasonably well isolated objects. In total, there are 552 IRAC sources
detected within the solid angle covered by the ACS images of the HUDF
(10.34 arcmin$^2$ in size). 

    We used the HUDF $z_{850}$-band-based catalog of Yan \& Windhorst (2004)
for optical identification. The magnitudes in this catalog are matched-aperture
``MAG\_AUTO'' magnitudes. The source matching was done by identifying the
closest ACS object within a $1''$ radius from the IRAC source centroid. The
$3.6\mu m$-to-$z_{850}$ flux density ratio
($\Gamma(3.6/z)\equiv f_\nu(3.6\mu m)/f_\nu(z_{850})$) histogram of the matched
sources has its peak at 1.7. We define IEROs as objects that have
$\Gamma(3.6/z) > 20$.
The unmatched IRAC sources automatically satisfy this criterion. No attempt has
been made to correct the differences in flux measurement that caused by the
different apertures used in generating the IRAC catalog and the $z_{850}$-band
catalog, since such differences are at most $\sim$ 0.1 mag (based on Monte
Carlo simulations using artificial sources and tests using larger photometry
apertures) and thus will not affect the selection for these extreme objects. 

   The major difficulty for reliable IERO selection and photometry is 
contamination due to source blending. Here, we adopt a conservative approach,
by visually inspecting each source in both the IRAC and the ACS + NICMOS HUDF
images, and considering only those sufficiently isolated objects that crowding
should not significantly affect the photometry. We inspected 75 IRAC sources
that met the criterion $\Gamma(3.6/z) > 20$, and rejected 58 that had nearby
objects that might have influenced the IRAC photometry, or that were clearly
the blended emission from two or more sources as revealed by the ACS images.
The remaining 17 objects constitute our final sample. Fifteen of
these objects do not have any companion within a $1''$ radius from the IRAC 
centroids as seen in the ACS images. The other two have
one or two close neighbors in the ACS images. However, we believe that their
identifications are secure for two reasons: in the ACS bands these neighbors
are all at least 1.0 mag fainter than the identified counterparts, and are all
centered more than $0.5''$ from the IERO centroids. 

  Table 1 gives the coordinates and photometric properties of these 17 IEROs.
The source fluxes are in the range
$0.64 < f_\nu(3.6\mu{\rm m})/\mu{\rm Jy} < 11$.
Among these objects, 12 are within the field coverage of the HUDF NICMOS 
Treasury Program (Thompson et al. 2004, in preparation), of which 11
(including the two objects
that have no optical counterparts) are identified in both the $J_{110}$ and the
$H_{160}$ bands. The unidentified one is the faintest IRAC source in our sample
(object \#10).  In addition, we also searched for their counterparts in the
deep $K_s$-band images obtained by ISAAC at VLT, which were taken as part
of the ground-based supporting data for GOODS (Giavalisco et al. 2004). Fifteen
of the IEROs have counterparts in the $K_s$ images, and the other two are not
seen because they are too faint. The $J_{110}$, $H_{160}$ and $K_s$ magnitudes
of the IEROs are also listed in Table 1 when available. As an example, the
image cut-outs of one of these objects are displayed in Figure 1.

\section{The Nature of IEROs}

    Without exception, the IEROs that have optical counterparts
all show a monotonic trend of increasing in flux from \iband\, to \chone-band,
and most of them show this trend starting in the \bband. This flux-increasing
trend closely resembles that of the ``unusual infrared object'' found by
Dickinson et al. (2000) in the HDF-North, whose nature remains uncertain.

   Here, we look for the simplest explanation for the IERO population as a
whole. The red nature of these objects immediately leads to a number of
broad possibilities: Galactic brown dwarfs, old and/or dusty galaxies, and
objects at extremely high redshifts. We can rule out the possibility that they
are stars in the Galaxy, because they are all resolved in the HST images
(ACS and/or NICMOS). Furthermore, as 15 out of the 17 objects
are visible in the ultra-deep optical images to at least \iband,
we can also reject the very-high-z ($z>7$) interpretation. In fact, most of
these objects are visible even in \bband, indicating that they reside
at $z\lesssim 4$.

    On the other hand, the IR parts of their SEDs seem to be consistent with
those of evolved stellar populations. Indeed, a SED template of local E/S0
(e.g., Coleman, Wu \& Weedman 1980) redshifted to $z\simeq$
1--3 can reasonably fit most of the objects in the IR region. Such an empirical
template, however, has a major drawback that its age ($>11$ Gyr) is much older
than the age of the universe at the inferred redshifts. To overcome this
problem, we explored the stellar population synthesis models of 
Bruzual \& Charlot (2003), which provides the flexibility of adjusting the
parameters of a model galaxy, especially its age and star-formation history. 
We used a Chabrier IMF (Chabrier 2003) and solar metallicity.
As an extremely red color can also be produced by dust, we also considered the
effect of dust using the extinction laws of Calzetti et al. (2000; at
$\lambda < 2000$\AA) and Fitzpatrick (1999; at $\lambda > 2000$\AA). 

    For most of the IEROs, we find that the shape of their SEDs in the 1 to 8 
$\mu$m wavelength range is best explained by the presence of a well-evolved
stellar component with ages in the range of 1.5--3.5 Gyr, observed at redshifts
$1.6 < z < 2.9$.  The gentle curvature of the infrared SED is like that 
expected from older stars, whose $f_\nu$ flux density peaks around 1.6$\mu$m in
the rest frame (e.g., Simpson \& Eisenhardt 1999). It is not well matched by
models dominated by heavily dust-obscured, young starlight (Fig. 2, top panel).
Although 1.5--3.5 Gyr-old stars formed in an instantaneous burst (or Simple
Stellar Population; SSP) match the infrared SED well, such a model 
significantly under-predicts the observed optical (rest-frame ultraviolet) 
fluxes of these objects. We find that
this can be solved by adding a weaker component of younger stars, which we 
model as a secondary instantaneous burst with an age of $\sim 0.1$~Gyr (Fig. 2,
bottom panel).  This two-component scheme can explain the majority of the IEROs
if the composite spectra are put at the right redshifts. The best fit for each
object requires fine-tuning the ages of both components, but this level of
adjustment is not well motivated here, since other variables may also be
important, including more complex past star formation histories, metallicity,
and dust extinction. For the sake of simplicity, we choose to fit all 17 IEROs
by the combination of a 2.5 and a 0.1 Gyr-old bursts, leaving only the 
redshift and the burst amplitudes as adjustable parameters. We emphasize that
this two-component model is a highly simplified approximation to the actual
physical processes and should not be over-interpreted. In particular, this
model does not exclude the possibility that dust reddening may play a role in
the galaxy colors, but simply emphasizes that the SED shape strongly favors the
presence of a dominant older stellar population, formed at a significantly
higher redshift than that at which the galaxy is observed. The young, secondary
component represents weaker star formation that was on-going or recently 
completed at the observed redshift. The two-component model generally provides
a much better fit than does the young, dusty model. As an example, the 
goodness-of-fit for any of the top panel models in Fig. 2 is 3--4$\times$ worse
than the two-component fit shown in the bottom. The GOODS MIPS 24$\mu$m 
observations will further help distinguish if any of these IEROs are dusty. For
instance, if they have only an insignificant amount of dust, our two-component
model predicts that the brightest IEROs have 24$\mu$m flux densities only at 
$\sim$ 2$\mu$Jy level and thus will not be seen even by the deep GOODS
24$\mu$m observations. On the other hand, if the IEROs are very dusty, their 
24$\mu$m flux will be much higher and thus could possibly be detectable.

   We also note that there is a degeneracy between the photometric redshift 
obtained during the fit and the ages of the components (especially the
primary components). Choosing an instantaneous burst older than 2.5 Gyr as the 
primary component does not change $z_p$ significantly (at a level of 
$\Delta z\sim 0.1$), but using a burst younger than 2.5 Gyr does tend to
increase the derived $z_p$, with $\Delta z\sim 0.4$ in extreme cases.
However, the overall $z_p$ values are not likely much
different from the current fits, since the plausible templates can neither be
much older than 2.5 Gyr because of the constraint from the age of the universe,
nor much younger than 1.5 Gyr because of the shape of the SEDs.

   Fig. 3 shows the SEDs of all 17 IEROs (filled boxes), along with the
fitted two-component models (open boxes connected by dashed lines). Four of
these objects have very uncertain photometric redshifts. Object \#10 
(the faintest in the sample) is only significantly detected in four of the
bands considered here, and thus can be fitted by templates at a variety of
redshifts. Object \#1 and \#2 do not have optical detections and can be fitted
by templates over a large redshift range. They are noteworthy, however, as
their estimated redshifts (3.6 and 3.4, respectively, albeit very uncertain),
are the highest among the IEROs. The SED for object \#5 is different
from that of any other object in the sample, with steeply increasing flux
density from 3.6 to 8$\mu$m, and yields the worst fit from our chosen models.
Therefore, these four objects are excluded from most of the discussion below.
The remaining 13 objects have $z_p$ values ranging
from 1.6 to 2.9, with the median at $z_p=2.4$. The adopted model template
is self-consistent with the redshift lower bound of $z=1.6$, because it
gives $\Gamma(3.6/z)<16$ at $z<1.6$, which does not meet our color selection
criterion even considering systematic errors in photometry. We refer to the
sample of these 13 objects as the refined sample. Their $z_p$ values are listed
in Table 2. This table also lists the corresponding rest-frame $K_s$ absolute
magnitudes of these objects, which are derived using the \chone\ measurements
and the K-corrections based on the fitted model SED templates. The masses of
the primary, older stellar components in the refined sample range from 0.1 to
1.6$\times 10^{11} M_\odot$ (with a median value of 5$\times 10^{10} M_\odot$),
which are 0.1 to 2$\times$ the characteristic $M^*=8.3\times 10^{10} M_\odot$
at $z=0$ from Cole et al. (2001; rescaled to account for the choice of the
Chabrier IMF that we adopt here). Using a younger template, 
for example, a 1.5 Gyr-old template, will decrease the mass estimate by no more
than a factor of two. The masses of the secondary bursts are 2--3 dex less. 
The median estimated masses of these objects is
$\sim 9\times$ larger than the typical ``best fit'' stellar masses
estimated by Papovich et al. (2001) for $L^*$ Lyman break galaxies (LBGs) at
$z \simeq 3$ ($\sim 6\times 10^9 M_\odot$, again adjusting for the
choice of IMF) and $\sim 3\times$ larger than the LBG masses derived by those
authors using their ``maximum M/L'' models. 

    Two of these IEROs, \#7 and \#9, are detected in X-ray by {\it Chandra}
(Giacconi et al. 2001; XID 515 and 605, respectively). The X-ray hardness 
ratios of both sources are very high, indicative of heavily absorbed emission.
Using the derived $z_p$ and assuming an intrinsic X-ray power-law slope of
slope of $\Gamma=1.8$, we estimate the obscuring column densities and the
unobscured X-ray luminosities ($L_X$) in rest-frame 0.5--7keV. Source \#7 has 
$L_X$ $\simeq 1.1\times 10^{44}$ erg\ s$^{-1}$ with
$\log(N_H) \simeq 23.5$; for \#9, which is detected solely in the hard 2--7 keV
band, we obtain $L_X > 2\times 10^{43}$ erg\ s$^{-1}$, assuming a $1\sigma$
lower limit on intrinsic absorption $\log(N_H) \gtrsim 23.3$. With X-ray
luminosities far exceeding that of starburst galaxies ($L_X \lesssim 10^{42}$
erg\ s$^{-1}$), these sources are likely obscured (Type 2) AGN, but are below
the QSO regime ($L_X \gtrsim 10^{44.5}$ erg\ s$^{-1}$; see Norman et al. 2002).
We estimate their observer-frame \chone\ AGN emission by extrapolating from the
unobscured $L_X$ assuming a flat $\nu F_\nu$ (e.g., Elvis et al. 1994). Even
assuming negligible rest-frame NIR extinction, the predicted \chone\ 
contribution from the AGN would be 0.575 and 0.09 \ujy\ for \#7 and \#9,
respectively, or 11\% and 1.5\% of the measured IERO flux densities. We 
therefore conclude that the AGN contribution to their optical-infrared SED is
safely ignored and the fitting described above is not affected by the AGN.

\section{Discussion}
  
   It is instructive to consider the relation between IEROs and other classes
of faint, red galaxies, such as conventional EROs. From the $K_s$ magnitudes
and the interpolated $R$ magnitudes listed in Table 1, one can see that nearly
all IEROs do satisfy the conventional ERO criterion of $(R-K)_{AB}>3.35$ mag 
(or $(R-K)_{Vega}>5$ mag). Are IEROs just EROs, but selected at redder 
wavelengths? The answer does not seem to be that simple, because we find that
most bright EROs do not meet the IERO color criterion. Moustakas et al. (2004)
present a sample of EROs in the CDF-S with $K_s < 22$ mag, 16 of which are in
the HUDF and match IRAC sources. However, none of these 16 EROs is selected as
an IERO, either because they do not pass the color criterion or because their
IRAC photometry is uncertain due to blending. As all but one of our IEROs have
$K_s > 22$ mag, it is important to consider fainter ERO samples. We use 
interpolated $R$-band magnitudes and the ISAAC $K_s$-band data to select a
sample of 27 fainter EROs at $22<K_s<24$ mag. Two-thirds (18/27)
of them match the initial list of 75 IERO candidates, including sources that
we the rejected due to blending issues. Among these 18 fainter EROs, 13 
match our IERO sample, i.e., roughly 50\% (13/27) of this fainter sample of
EROs qualify as IEROs. However, another 33\% (9/27) do not match any IERO
candidates.

   Therefore, while most bright ($K_s<22$ mag) EROs are not IEROs, there is
a substantial (but not complete) overlap between the two selection criteria at
fainter magnitudes. A consistent interpretation would be that the IERO color
criterion picks up the fainter, higher redshift equivalents of conventional
EROs.  The photometric redshifts of our IEROs are at $1.6\leq z\leq 2.9$
(possibly extending to $z < 3.6$), with the median at $z\simeq 2.4$. By
contrast, redshifts (photometric and spectroscopic) for the Moustakas et al.\
ERO sample peak at $z\simeq 1$ (only a minority at $1.6<z<2.5$), with the 
median at $z\simeq 1.2$. Furthermore, the ($z_{850}$, 3.6$\mu$m) filter pair
samples roughly the same rest-frame wavelength range at $z\simeq 2.4$ as does
the ($R$, $K_s$) filter pair at $z\simeq 1.2$, and $\Gamma(3.6/z)>20$ 
corresponds to $(z_{850}-m_{3.6})>3.25$ mag, i.e., the IERO criterion selects
roughly the the same rest-frame feature at higher redshifts as the ERO 
criterion does at lower redshifts.

   Several authors (e.g., Totani et al.\ 2001; Franx et al.\ 2003; Saracco
et al.\ 2004; Chen \& Marzke 2004) have used near-infrared colors to identify
high redshift galaxy candidates. All of our IEROs with sufficiently deep 
$J$-- and $K$--band photometry satisfy the $(J-K)_{Vega}>2.3$ mag criterion
used by Franx et al., and have a similar surface density on the sky. Five
IEROs also satisfy the $(J-K)_{Vega}>3$ mag criterion used by Totani et al.\
and Saracco et al. Most of these objects populate the more distant end of our
photometric redshift range ($2.7 \leq z_p \leq 3.6$). Chen \& Marzke (2004)
report nine faint, red galaxies selected from the HUDF ACS and NICMOS data,
with $(i_{775}-H_{160})>2$ mag and $z_p>2.5$. Four of our IEROs (\#1, 2, 8 and
9) are also in the Chen \& Marzke list. Most of these authors have suggested
that these red, IR-selected galaxies have relatively massive, old stellar
populations at $z>2$, although Totani et al.\ interpret them as highly 
obscured, star-bursting proto-ellipticals. By extending photometry to 8$\mu$m 
with IRAC, we also find that the SED shapes for most of the IEROs favor the
presence of dominant, old stellar populations.

   If most of these IEROs are indeed at $z\simeq 1.6$--2.9, their 
IR luminosities indicate that they are at the bright end of the
luminosity function. The rest-frame $K_s$-band absolute magnitudes of the
objects in the refined sample range from $M_{AB}(K_s)=-21.3$ to $-24.2$
mag, while the median is at $-22.9$ mag. Most of these objects are 
significantly brighter than the near-IR $M^*$ value in the local universe. For
example, the 2MASS $K_s$-band LF of Kochanek et al. (2001) gives 
$M^*(K_s)=-23.53+5\log(h)$ for early type galaxies, which corresponds to 
$M_{AB}^*(K_s)=-22.43$ mag.

   A detailed treatment of the space density of IEROs is beyond the scope of
this paper, and requires a more extensive analysis of source detection
efficiency and incompleteness in the crowded IRAC images, which is presently
underway. Here, we limit ourselves to some simple comparisons to the number
density of local, early--type galaxies, which we regard as likely descendants
of the IEROs. We assume the volume over which IEROs can be observed extends
from $z = 1.6$, consistent with the lowest redshift at which an unreddened old
stellar population should meet the IERO color criterion, to $z = 2.9$, the
highest redshift for an object in the refined sample. The least luminous
IERO (\#12) in the refined sample has $M_{AB}(K_s) = -21.39$ mag at $z_p = 1.9$.
If we assume a roughly flat SED (in $f_\nu$ vs. $\lambda$) in the IRAC bands
(as is typically observed), and redshift it to $z = 2.9$, this galaxy would
have 
$m(3.6\mu{\rm m}) = 24.1$ mag, and would be easily detected with a formal point
source S/N$\sim$38. The actual detection likelihood will be smaller due to the
blending, which affects our catalogs over a wide range of flux. We neglect this
here, and naively assume that the refined sample represents a lower bound to a
complete and volume-limited set of objects with $M_{AB}(K_s) \leq -21.3$ mag
at $1.6\leq z\leq 2.9$.
 
   Within this volume, the refined IERO sample with $M_{AB} < -21.3$ mag
has a space density of $3.2 \times 10^{-4}$~Mpc$^{-3}$, of which the six most
luminous objects (with $M_{AB}(K_s) < -23.0$ mag; hereafter ``the luminous
subsample'') contribute $1.3 \times 10^{-4}$~Mpc$^{-3}$. The actual number
density of such objects is certainly larger, due to catalog incompleteness and
the conservative screening process during the selection. We compare these
number densities to those calculated by integrating the local $K_s$--band LF
for early--type galaxies (Kochanek et al. 2001). For the (unphysical) case of no
evolution (NE), the local LF predicts 12.4 and $1.3 \times 10^{-4}$ objects per
Mpc$^3$ for $M_{AB}(K_s)\leq -21.3$ and $\leq -23.0$ mag, respectively. The
(likely incomplete) IERO sample therefore makes up 25\% of the number density
of early--type galaxies with comparable luminosities in the local universe,
while the luminous subsample has virtually the same space density as for
comparably luminous local ellipticals. We might assume instead that the IERO
population undergoes pure luminosity evolution (PLE) down to the present day.
A simple, single--age stellar population fades by 1.0 mag in the rest--frame
$K$--band from an age of 2.5 to 13 Gyr (Bruzual \& Charlot 2003). In this
scenario, the entire refined sample and the luminous subsample make up 
14\% and 20\% of the corresponding local number densities of early--type
galaxies with $M_{AB}(K_s)\leq -20.3$ and $\leq -22.0$ mag, respectively.
We note that the galaxies in the luminous subsample are found entirely at the
upper end of the redshift range, $2.4\leq z \leq 2.9$, and if this volume were
used instead, their space densities would be 2.55$\times$ larger, making up
51\% of the local population under the PLE model to $M_{AB}(K_s)\leq -22.0$ mag.
 
   The HUDF samples a small cosmic volume, and we may expect the most luminous
and massive galaxies to cluster strongly, potentially leading to large 
field--to--field variance in the densities. We must therefore regard these 
simple estimates with caution, pending analysis of the full GOODS area, with
proper treatment of sample incompleteness (although we note that only the HUDF
has sufficiently deep optical imaging data to enable firm color constraints on
IEROs down to the flux limit of the {\it Spitzer} IRAC images). Nevertheless,
we may say that the IEROs appear to be a significant and numerous population of 
objects, most naturally interpreted as galaxies with relatively old and massive
stellar populations at $z > 1.6$. 

   To summarize, using the first epoch of the GOODS {\it Spitzer} Legacy
Program observations in the CDF-S, we have identified 17 well-isolated objects
in the HUDF that are significantly detected by IRAC but are very faint in the
ACS images. Their SEDs are best explained by the presence of an old 
($\sim 1.5$ to 2.5 Gyr) stellar population in galaxies at $1.6 < z < 2.9$. A
few of the objects may have higher redshifts, but this is yet uncertain. The
old stars dominate the infrared light, with a median rest-frame $K_s$-band
luminosity that is $1.5\times$ that of present--day $L^\ast$ early-type
galaxies, and stellar masses $\sim 0.1$ to $1.6\times 10^{11} M_\odot$ for a 
Chabrier IMF. They are substantially more massive than typical Lyman break
galaxies at similar redshifts. A much smaller component of recent star 
formation is needed to explain the optical (rest-frame UV) portion of the SED.
The IEROs are likely the higher-redshift analogs of conventional EROs, selected
via similar color criteria applied at longer wavelengths. The GOODS IRAC 
observations contribute to the mounting evidence for a significant population
of red, evolved galaxies at high redshifts. In a simple PLE scenario, the IEROs
may be direct progenitors for at least 14 to 51\% of the local population of
massive, early-type galaxies.

\acknowledgments

   Support for this work, part of the {\it Spitzer Space Telescope}
Legacy Science Program, was provided by NASA through Contract Number 1224666
issued by the Jet Propulsion Laboratory, California Institute of Technology
under NASA contract 1407.

\begin{figure}
\plotone{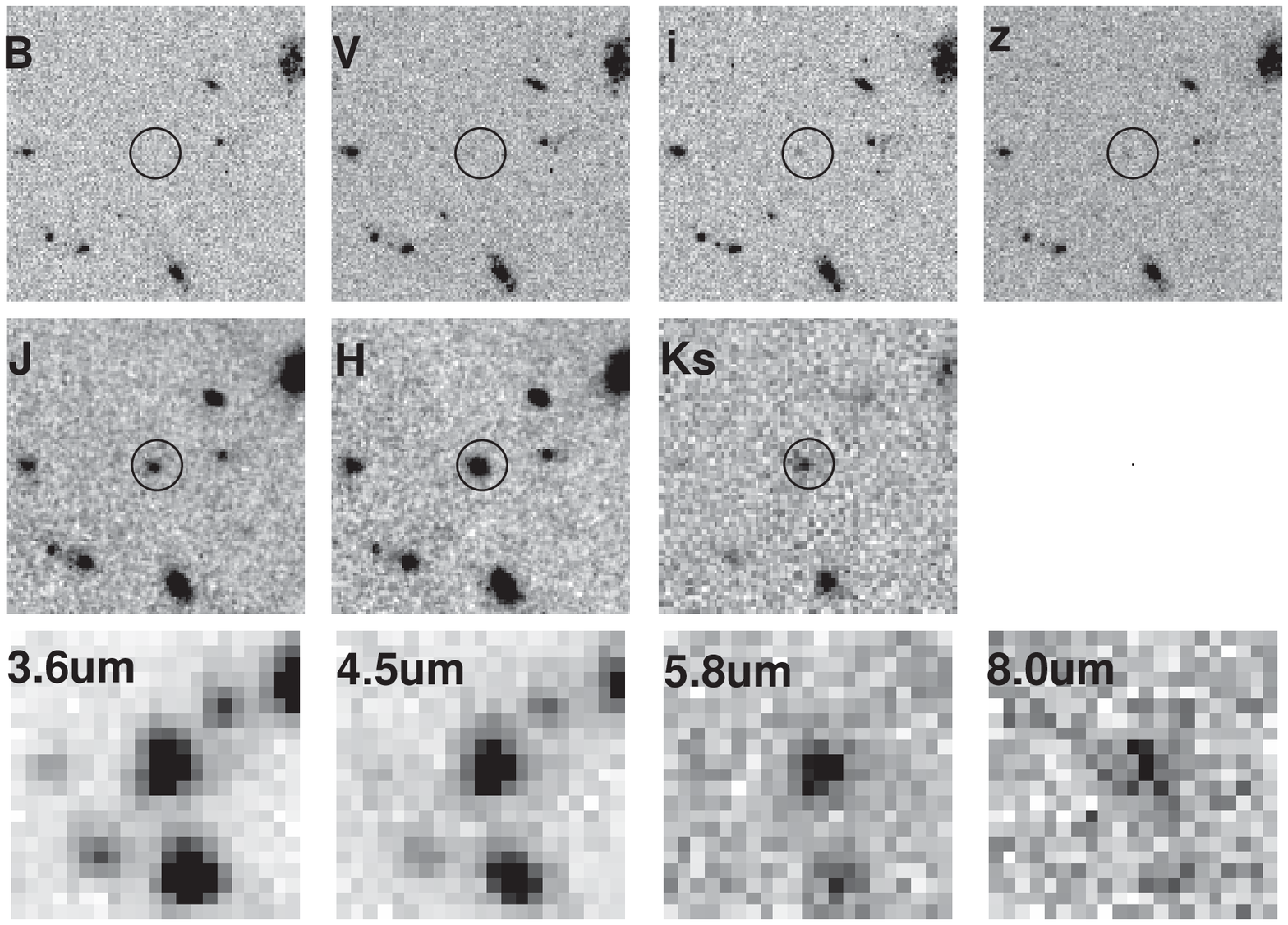}
\caption{As an example of IRAC-selected extremely red objects, 11-band image
cut-outs of object \#4 in Table 1 are shown here. Images are 
12$^{''}\times$12$^{''}$ with N up and E to the left. The optical (top panel)
$BViz$ images are from the HUDF ACS campaign. The near-infrared (middle panel)
$JH$ images are from the HUDF NICMOS Treasury Program, while the $Ks$ image
is from the data obtained at the VLT/ISAAC as part of the GOODS ground-based
supporting observations. The IRAC images (bottom panel) are described in this
paper. The location of this source, as derived based on the IRAC \chone\ image,
is illustrated with a 1$^{''}$ radius circle in the top and the middle panels.
}
\end{figure}

\begin{figure}
\epsscale{0.9}
\plotone{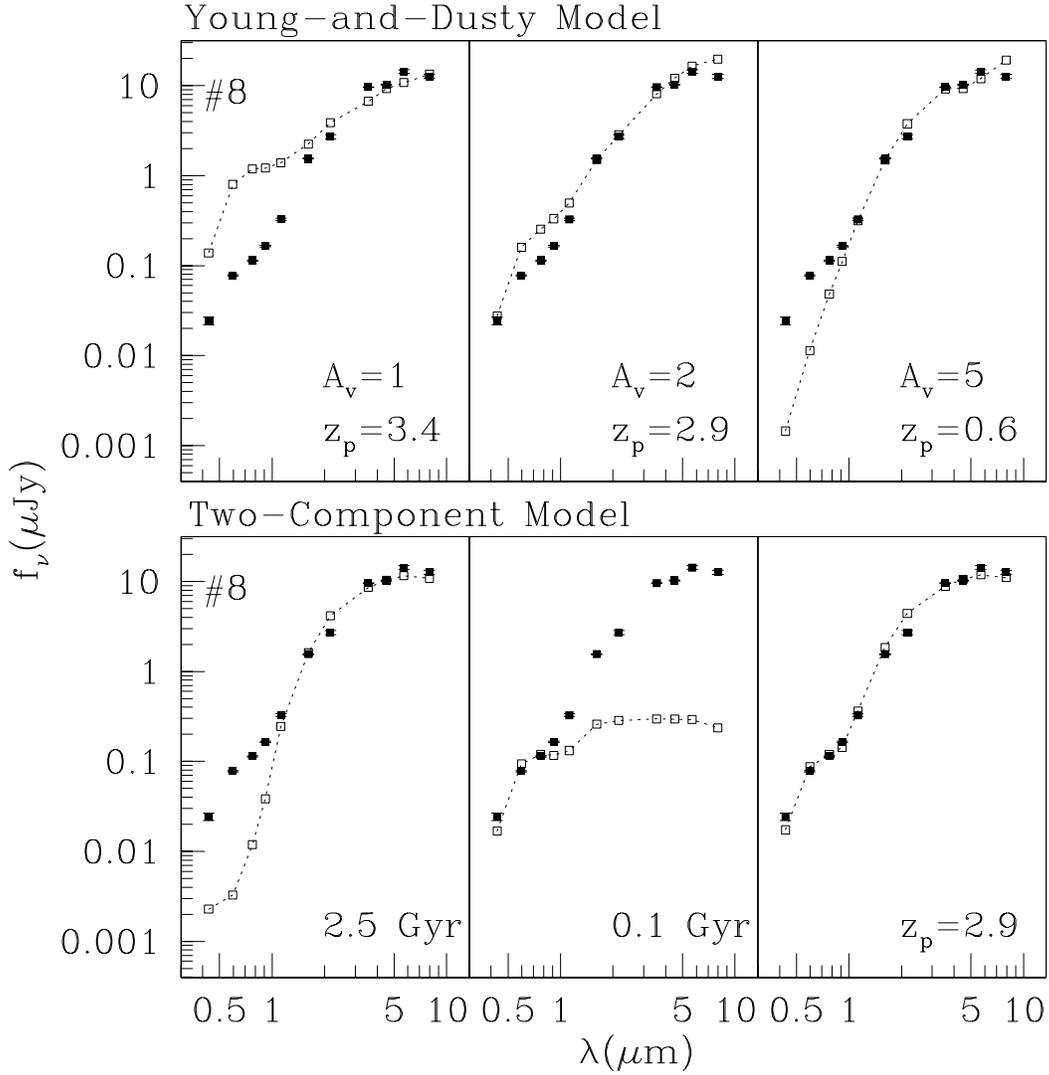}
\caption{({\it Top}) The SEDs of the IEROs cannot be easily explained without
a well-evolved stellar population.
The case for \#8 is demonstrated here. The observed SED is shown as filled
boxes while the template is shown as open boxes connected by dashed lines.
The template is a 0.1 Gyr-old star-bursting galaxy, reddened by dust
extinction of $A_V=1.0$, 2.0 and 5.0 mag for the left, middle and right panels,
respectively. The corresponding best-fit $z_p$ values are also labeled.
({\it Bottom}) The SEDs of most of the IEROs can be satisfactorily
explained by a two-component model, and the fitting is shown for the same object
\#8. The primary and the secondary components of the model are shown in the
left and the middle panels, respectively. The primary component is a massive, 
old stellar population approximated by a 2.5 Gyr-old instantaneous burst, while
the secondary component is a weak, on-going star-forming population
approximated by a 0.1 Gyr-old instantaneous burst. The right panel shows the
composition of these two components. The best-fit photometric redshift is
$z_p=2.9$.}
\end{figure}

\begin{figure}
\epsscale{0.9}
\plotone{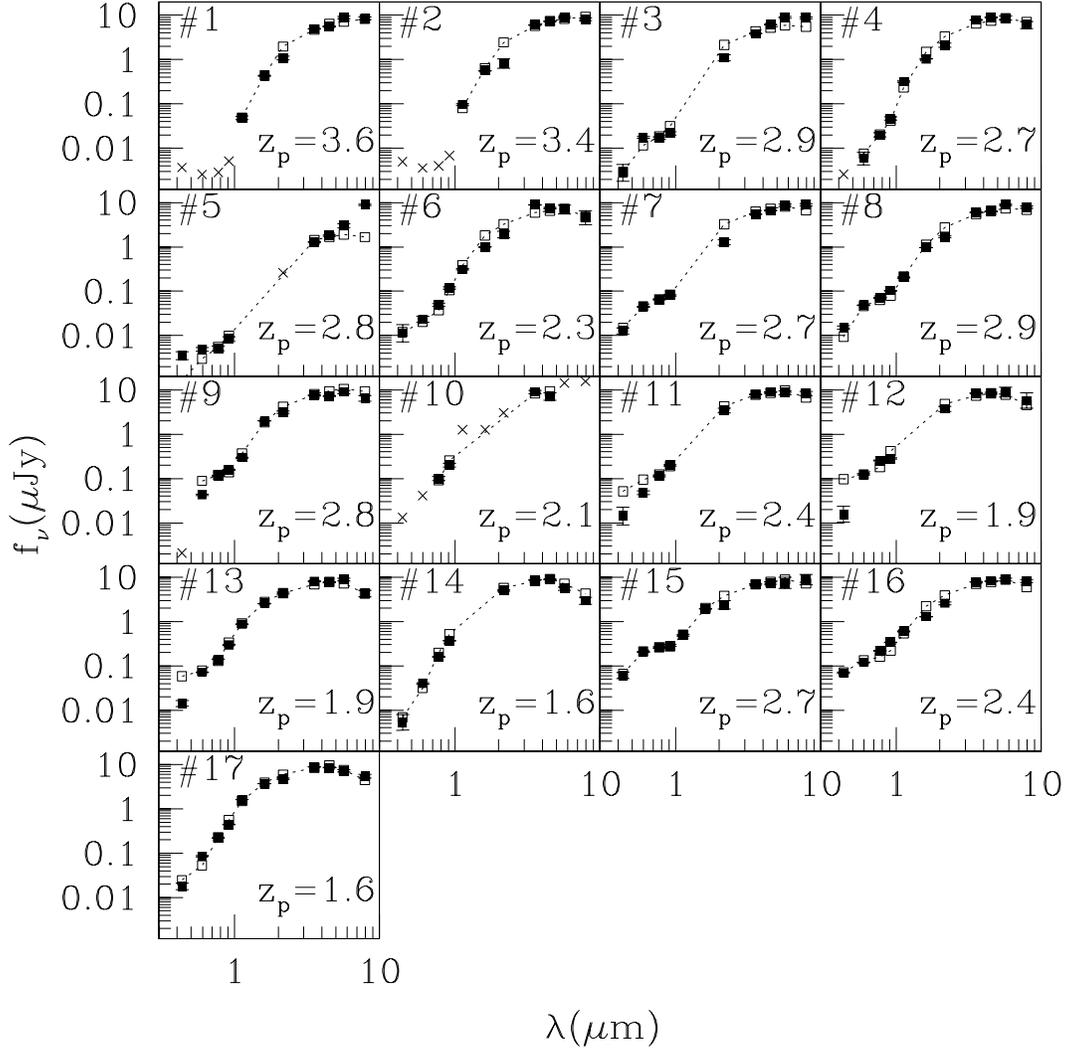}
\caption{For simplicity, a composite model of a 2.5 Gyr-old burst and a 0.1
Gyr-old burst is used to fit all the 17 objects. Legends are the same as in
Fig. 2. Flux density upper limits, when available, are shown as crosses. For
clarity, the SEDs of these objects have been brought to the same
scale. The objects in the refined sample (see text) have $z_p$ range from 1.6
to 2.9, with the median at 2.4. The median rest-frame $K_s$-band absolute
magnitude thus derived is --22.9 mag, which is significantly brighter than
the $M_{AB}^*(K_s)$ value of the local elliptical galaxy LF.}
\end{figure}

\begin{deluxetable}{rccccccccccccccc}
\tablewidth{0pt}
\tabletypesize{\tiny}
\rotate
\tablecaption{Photometric properties of IEROs selected in the HUDF}
\tablehead{
\colhead{ID} &
\colhead{RA\&DEC(J2000)} &
\colhead{$B_{435}$} &
\colhead{$V_{606}$} &
\colhead{$(R)$} &
\colhead{$i_{775}$} &
\colhead{$z_{850}$} &
\colhead{$J_{110}$} &
\colhead{$H_{160}$} &
\colhead{$K_s$} &
\colhead{m($3.6\mu m$)} &
\colhead{m($4.5\mu m$)} &
\colhead{m($5.8\mu m$)} &
\colhead{m($8.0\mu m$)} &
\colhead{$\Gamma(3.6/z)$}
}
\startdata

 1 & 3:32:42.86 -27:48:09.29&    $>29.8$    &  $>30.2$   &  ---  &  $>30.1$    &   $>29.5$     &27.01$\pm$0.10& 24.63$\pm$0.02& 23.63$\pm$0.09& 21.99$\pm$0.01 & 21.86$\pm$0.02 & 21.35$\pm$0.07 & 21.39$\pm$0.09 &  $>1009$ \\
 2 & 3:32:38.74 -27:48:39.96&    $>29.8$    &  $>30.2$   &  ---  &  $>30.1$    &   $>29.4$     &26.58$\pm$0.05& 24.68$\pm$0.01& 24.28$\pm$0.24& 22.10$\pm$0.01 & 21.88$\pm$0.01 & 21.67$\pm$0.05 & 21.78$\pm$0.07 &  $>831$ \\
 3 & 3:32:43.52 -27:46:38.96& 30.14$\pm$0.48& 28.14$\pm$0.06&(28.15)&28.18$\pm$0.07& 27.91$\pm$0.09&      ---     &       ---     & 23.65$\pm$0.18& 22.28$\pm$0.01 & 21.79$\pm$0.01 & 21.38$\pm$0.05 & 21.41$\pm$0.05 & 178.8 \\
 4 & 3:32:41.74 -27:48:24.84&    $>31.1$     & 30.22$\pm$0.36&(29.54)&28.88$\pm$0.12& 27.96$\pm$0.09&25.88$\pm$0.03& 24.60$\pm$0.01& 23.82$\pm$0.11& 22.40$\pm$0.01 & 22.26$\pm$0.02 & 22.30$\pm$0.12 & 22.69$\pm$0.20 & 167.3 \\
 5 & 3:32:43.66 -27:48:50.69& 29.71$\pm$0.23& 29.33$\pm$0.11&(29.32)&29.29$\pm$0.12& 28.76$\pm$0.13&      ---     &       ---     &       $>25.0$     & 23.28$\pm$0.03 & 22.92$\pm$0.04 & 22.37$\pm$0.11 & 21.19$\pm$0.05 & 156.6 \\
 6 & 3:32:33.25 -27:47:52.19& 29.75$\pm$0.50& 28.96$\pm$0.17&(28.60)&28.15$\pm$0.09& 27.20$\pm$0.06&26.14$\pm$0.05& 24.88$\pm$0.02& 24.17$\pm$0.18& 22.50$\pm$0.02 & 22.69$\pm$0.04 & 22.77$\pm$0.24 & 23.22$\pm$0.40 &  75.4 \\
 7 & 3:32:32.16 -27:46:51.60& 28.72$\pm$0.16& 27.36$\pm$0.03&(27.19)&26.94$\pm$0.03& 26.71$\pm$0.04&      ---     &       ---     & 23.73$\pm$0.15& 22.15$\pm$0.01 & 21.93$\pm$0.02 & 21.63$\pm$0.07 & 21.61$\pm$0.07 &  67.1 \\
 8 & 3:32:35.06 -27:46:47.46& 27.94$\pm$0.11& 26.67$\pm$0.02&(26.51)&26.26$\pm$0.02& 25.86$\pm$0.02&25.11$\pm$0.04& 23.42$\pm$0.01& 22.82$\pm$0.06& 21.44$\pm$0.01 & 21.37$\pm$0.01 & 21.01$\pm$0.05 & 21.14$\pm$0.06 &  58.6 \\
 9 & 3:32:39.16 -27:48:32.44&    $>30.8$      & 27.52$\pm$0.04&(27.01)&26.44$\pm$0.02& 26.12$\pm$0.02&25.45$\pm$0.03& 23.36$\pm$0.01& 22.87$\pm$0.06& 21.93$\pm$0.01 & 21.95$\pm$0.02 & 21.74$\pm$0.10 & 22.10$\pm$0.16 &  47.6 \\
10 & 3:32:37.15 -27:48:23.54&    $>31.4$     &   $>30.2$   &(30.38)&29.26$\pm$0.12& 28.49$\pm$0.10&   $>26.5$ &   $>26.5$ &       $>25.5$     & 24.38$\pm$0.09 & 24.62$\pm$0.21 &   $>23.9$ &   $>23.8$ &  44.3 \\
11 & 3:32:30.26 -27:47:58.24& 29.81$\pm$0.50& 28.50$\pm$0.10&(28.06)&27.55$\pm$0.05& 26.96$\pm$0.05&      ---     &       ---     & 23.87$\pm$0.14& 22.96$\pm$0.01 & 22.83$\pm$0.02 & 22.86$\pm$0.10 & 22.91$\pm$0.12 &  39.8 \\
12 & 3:32:29.82 -27:47:43.30& 30.11$\pm$0.45& 27.87$\pm$0.04&(27.53)&27.10$\pm$0.02& 26.98$\pm$0.03&      ---     &       ---     & 24.18$\pm$0.14& 23.30$\pm$0.03 & 23.31$\pm$0.05 & 23.24$\pm$0.27 & 23.71$\pm$0.44 &  29.4 \\
13 & 3:32:39.11 -27:47:51.61& 28.87$\pm$0.16& 27.11$\pm$0.02&(26.79)&26.37$\pm$0.01& 25.57$\pm$0.01&24.40$\pm$0.01& 23.23$\pm$0.01& 22.64$\pm$0.05& 21.98$\pm$0.01 & 22.03$\pm$0.02 & 21.87$\pm$0.08 & 22.68$\pm$0.18 &  27.2 \\
14 & 3:32:48.55 -27:47:07.58& 29.32$\pm$0.45& 27.12$\pm$0.04&(26.35)&25.63$\pm$0.01& 24.73$\pm$0.01&      ---     &       ---     & 21.89$\pm$0.04& 21.30$\pm$0.01 & 21.27$\pm$0.01 & 21.79$\pm$0.10 & 22.47$\pm$0.20 &  23.6 \\
15 & 3:32:38.76 -27:48:27.07& 28.54$\pm$0.18& 27.21$\pm$0.04&(27.12)&26.97$\pm$0.03& 26.84$\pm$0.05&26.27$\pm$0.04& 24.79$\pm$0.02& 24.56$\pm$0.23& 23.41$\pm$0.04 & 23.34$\pm$0.06 & 23.32$\pm$0.31 & 23.12$\pm$0.29 &  23.6 \\
16 & 3:32:35.71 -27:46:38.96& 27.26$\pm$0.04& 26.68$\pm$0.02&(26.40)&26.02$\pm$0.01& 25.54$\pm$0.01&24.94$\pm$0.03& 24.10$\pm$0.02& 23.34$\pm$0.08& 22.14$\pm$0.01 & 22.10$\pm$0.01 & 22.00$\pm$0.08 & 22.08$\pm$0.08 &  23.0 \\
17 & 3:32:33.67 -27:47:51.04& 28.45$\pm$0.19& 26.76$\pm$0.03&(26.27)&25.72$\pm$0.01& 24.97$\pm$0.01&23.56$\pm$0.01& 22.67$\pm$0.00& 22.39$\pm$0.04& 21.71$\pm$0.01 & 21.76$\pm$0.01 & 21.95$\pm$0.07 & 22.24$\pm$0.10 &  20.1 \\

\enddata

\tablenotetext{1.} {The magnitudes are in AB system, which are related to flux
density $f_\nu$ (in erg\ s$^{-1}$\ cm$^{-2}$\ Hz$^{-1}$) by
$m=-2.5\times lg(f_\nu) - 48.60$.}
\tablenotetext{2.} {The reported photometric errors of the IRAC bands reflect the random errors only. Typical systematic errors in these bands are at $\lesssim 0.1$ mag level.}
\tablenotetext{3.} {The \jband\ and \hband\ magnitudes are adapted from the
catalog released with the NICMOS HUDF data, which are also ``MAG\_AUTO''
magnitudes extracted by SExtractor. Objects \# 3, 5, 7 11, 12, and 13 are
outside of the NICMOS HUDF field; \# 10 is within the NICMOS field but is not
detected in either band.}
\tablenotetext{4.} {The $K_s$-band magnitudes are based on the deep ISAAC
images obtained at the VLT, which are ``MAG\_AUTO'' magnitudes extracted by
SExtractor using the updated zeropoints.}
\tablenotetext{5.} {The $(R)$ values are not measured but obtained by
interpolating from $V_{606}$ and $i_{775}$ to 6500\AA, and are listed here
only for comparing against the conventional (R-K) ERO definition.}

\end{deluxetable}

\begin{deluxetable}{cccccc}
\tablewidth{0pt}
\tabletypesize{\scriptsize}
\tablecaption{Photometric redshifts and absolute magnitudes of IEROs}
\tablehead{
\colhead{ID} &
\colhead{m(3.6$\mu m$)} &
\colhead{$z_p$} &
\colhead{$M$} &
\colhead{K-correction} &
\colhead{$M(K_s)$}
}
\startdata

 $1^*$ & 21.99 & 3.6&  ---   & ---   & ---  \\
 $2^*$ & 22.10 & 3.4&  ---   & ---   & ---  \\
 3 & 22.28 & 2.9 & $-23.22$ & $-0.17$ & $-23.39$ \\
 4 & 22.40 & 2.7 & $-22.96$ & $-0.10$ & $-23.06$ \\
 $5^*$ & 23.28 & 2.8& ---   & ---   & ---  \\
 6 & 22.50 & 2.3 & $-22.56$ & $ 0.00$ & $-22.56$ \\
 7 & 22.15 & 2.7 & $-23.22$ & $-0.10$ & $-23.32$ \\
 8 & 21.44 & 2.9 & $-24.06$ & $-0.17$ & $-24.23$ \\
 9 & 21.93 & 2.8 & $-23.51$ & $-0.14$ & $-23.65$ \\
 $10^*$ & 24.38 & 2.1&  ---   & ---   & ---  \\
11 & 22.96 & 2.4 & $-22.19$ & $-0.02$ & $-22.21$ \\
12 & 23.30 & 1.9 & $-21.39$ & $ 0.06$ & $-21.33$ \\
13 & 21.98 & 1.9 & $-22.71$ & $ 0.06$ & $-22.65$ \\
14 & 21.30 & 1.6 & $-23.05$ & $ 0.11$ & $-22.94$ \\
15 & 23.41 & 2.7 & $-21.96$ & $-0.10$ & $-22.06$ \\
16 & 22.14 & 2.4 & $-23.01$ & $-0.02$ & $-23.03$ \\
17 & 21.71 & 1.6 & $-22.64$ & $ 0.11$ & $-22.53$ \\

\enddata

\tablenotetext{1.} {The four objects marked with asterisks have very uncertain
photometric redshifts and are not included in the refined sample.}
\tablenotetext{2.} {The $M$ values are the absolute magnitudes at the rest-frame
wavelength that correspond to the observer-frame \chone, while $M(K_s)$ values
are the absolute magnitudes at the rest-frame $K_s$-band obtained by adding the
K-corrections to the $M$. The K-corrections are calculated using the fitting
templates as shown in Fig. 3.}

\end{deluxetable}


\begin{thebibliography}{}

\bibitem[]{411} Bertin, E. \& Arnouts, S. 1996, A\&AS, 117, 393
\bibitem[]{412} Bruzual, A. G. \& Charlot, S. 2003, MNRAS, 344, 1000
\bibitem[]{413} Calzetti, D., et al. 2000, ApJ, 533, 682
\bibitem[]{} Chen, H.-W. \& Marzke, R. 2004, submitted to ApJL (astro-ph/0405432)
\bibitem[]{} Chabrier, G. 2003, PASP, 115, 763
\bibitem[]{414} Cole, S., et al. 2001, MNRAS, 326, 255
\bibitem[]{415} Coleman, G. D., Wu, C.-C., \& Weedman, D. W. 1980, ApJS, 43, 393
\bibitem[]{} Daddi, E., Cimatti, A., \& Renzini, A. 2000, A\&A, 362, L45
\bibitem[]{416} Dickinson, M., et al. 2000, ApJ, 531, 624
\bibitem[]{} Elstion, R., Rieke, G. H., \& Rieke, M. J. 1988, ApJ, 331, L77
\bibitem[]{418} Elvis, M., et al. 1994, ApJS, 95, 1
\bibitem[]{419} Fazio, G. G., et al. 2004, accepted for publication in ApJS (astro-ph/0405616)
\bibitem[]{421} Fitzpatrick, E. L. 1999, PASP, 111, 63
\bibitem[]{422} Franx, M., et al. 2003, ApJ, 587, L79
\bibitem[]{423} Giacconi, R., et al. 2001, ApJ, 551, 624
\bibitem[]{} Hu, E. M. \& Ridgway, S. E. 1994, AJ, 107, 1303
\bibitem[]{424} Kochanek, C. S., et al. 2001, ApJ, 560, 566
\bibitem[]{} McCarthy, P. J., Persson, S. E., \& West, S. C. 1992, ApJ, 386, 52
\bibitem[]{425} Moustakas, L. A., et al. 2004, ApJ, L131
\bibitem[]{426} Norman, C., et al. 2002, ApJ, 571, 218
\bibitem[]{427} Papovich, C., Dickinson, M \& Ferguson, H. C. 2001, ApJ, 559, 620
\bibitem[]{} Scodeggio, M. \& Silva, D. R. 2000, A\&A, 359, 953
\bibitem[]{} Simpson, C. \& Eisenhardt, P. 1999, PASP, 111, 691
\bibitem[]{} Thompson, D., et al. 1999, ApJ, 523, 100
\bibitem[]{428} Totani, T., Yoshii, Y., Iwamuro, F., Maihara, T., \& Motohara, K., 2001, ApJ, 558, L87
\bibitem[]{} Werner, M. W., et al. 2004, accepted for publication in ApJS (astro-ph/0406223)
\bibitem[]{429} Yan, H. \& Windhorst, R., 2004, accepted for publication in ApJL (astro-ph/0407493)
\bibitem[]{} Yan, L., McCarthy, P. J., Weymann, R. J., Malkan, M. A., Teplitz, H. I., Storrie-Lombardi, L. J., Smith, M., \& Dressler, A. 2000, AJ, 120, 575
\end{thebibliography}
\end{document}